\documentclass[journal]{IEEEtran}

\usepackage{amsmath,amssymb}
\usepackage{graphicx}
\usepackage{booktabs}
\usepackage{multirow}
\usepackage{algorithm}
\usepackage{algpseudocode}
\usepackage{cite}          % IEEE-standard numeric citation package (sorts/compresses automatically)
\usepackage[hidelinks]{hyperref}  % loaded after cite, matching IEEEtran's recommended package order

\title{A Controlled Evaluation of Quantum Correlation Refinement for Few-Shot Semantic Segmentation: Resource Cost and IBM Quantum Hardware Validation}

\author{Hina Shakir,%%
        Muhammad Irfan Memon,%
         Asia Samreen,%
        Javeed Hussain,%
        Syed Rizwan Ali,%
        Mohammad Mohatram,%
        and Muhammad Hussain
\thanks{H. Shakir is with the Department of Software Engineering, Bahria University, Karachi, Pakistan (e-mail: hinashakir.bukc@bahria.edu.pk).}
\thanks{A. Samreen is with the Department of Computer Science, Bahria University, Karachi, Pakistan (e-mail: asiasamreen.bukc@bahria.edu.pk).}
\thanks{M. Mohatram, J. Hussain, and M. I. Memon are with Global College of Engineering and Technology, Muscat, Oman (e-mail: m.mohatram@gcet.edu.om; s.javeedhussain@gcet.edu.om; m.memon@gcet.edu.om).}
\thanks{S. R. Ali is with the Software Engineering \& Business Incubation Center, Bahria University, Karachi, Pakistan (e-mail: rizwan257@gmail.com).}
\thanks{Corresponding authors: M. Mohatram, J. Hussain, and M. I. Memon (e-mail: m.mohatram@gcet.edu.om; s.javeedhussain@gcet.edu.om; m.memon@gcet.edu.om).}}
\begin{document}
\maketitle

\begin{abstract}
Parameterized quantum circuits (PQCs) are increasingly proposed as trainable components within classical machine-learning pipelines, but their resource cost, trainability, and hardware behavior are rarely characterized alongside a controlled measurement of task-level benefit. This paper reports such a controlled engineering evaluation, using few-shot semantic segmentation as a dense-prediction testbed. We integrate a Quantum Correlation Refiner (QCR) -- a six-qubit variational circuit with amplitude embedding and a gated residual connection -- into a classical correlation-based segmentation architecture, and evaluate it under the four-fold PASCAL-5$^i$ protocol with three seeds per fold, against an unrefined classical baseline and a parameter-matched classical MLP refiner. Across 12 matched fold--seed comparisons, QCR changes mean intersection-over-union (mIoU), the standard class-averaged segmentation accuracy metric, by only $+0.0001$ relative to the classical baseline ($t(11)=0.10$, $p=0.92$, $d_z=0.03$); the MLP control and an expanded Pauli-measurement variant show similarly no significant improvement. We characterize the module's engineering cost directly: it adds only 473 trainable parameters but roughly doubles per-epoch training time, scaling further with measurement dimensionality. We further validate a trained QCR circuit on IBM quantum hardware; for one checkpoint and 18 validation patches, hardware expectation values agree closely with the noiseless simulator ($r=0.954$, MAE $=0.101$), a circuit-fidelity check rather than an accuracy evaluation. Together, these results show the evaluated PQC module is fully functional end-to-end -- trainable, hardware-deployable, and numerically faithful to simulation -- while providing no measurable task-level advantage over a parameter-matched classical alternative under this architecture and resource regime. The study offers a template for evaluating small quantum components inside classical systems: parameter-matched controls, repeated seeds, resource measurement, and simulator-to-hardware validation applied jointly, rather than any one alone.
\end{abstract}


\IEEEpeerreviewmaketitle

\section{Introduction}
\label{sec:intro}
Few-shot semantic segmentation (FSS) aims to segment previously unseen object
categories using only a small number of annotated support examples. Unlike
conventional semantic segmentation, which typically relies on large collections
of pixel-level annotations, FSS transfers semantic information from the support
images to an unlabeled query image. This makes reliable support--query feature
matching particularly important for distinguishing the target foreground from
visually similar background regions~\cite{Shaban2017OSLSM,Zhang2018SGOne,Wang2019PANet}.

Recent FSS methods have increasingly adopted feature correlation and
prototype-based matching to establish dense correspondence between support and
query images. Methods such as PFENet and HSNet demonstrate the effectiveness of
dense similarity representations for few-shot segmentation~\cite{Zhang2021PFENet,Min2021HSNet}.
However, these representations remain dependent on classical feature
transformations and similarity operations, motivating investigation of alternative
computational mechanisms within established FSS pipelines.

Quantum machine learning provides one such possibility through parameterized
quantum circuits (PQCs), which can be integrated with classical neural networks
to form hybrid quantum--classical models~\cite{Cong2019QCNN,Henderson2020Quanvolutional}.
Parameterized circuits provide trainable quantum transformations that can be
embedded within classical learning pipelines, although their practical benefit
for computer-vision tasks remains an empirical question under current
resource and noise constraints.
In parallel, a growing body of work has explored variational quantum circuits as feature extractors, kernels, or refinement modules within otherwise classical deep-learning pipelines~\cite{Biamonte2017QML,Havlicek2019QuantumKernels,Henderson2020Quanvolutional,Mari2020TransferLearning,Cong2019QCNN}. Most of these studies, however, have been evaluated on small benchmarks, many of which are synthetic or closely aligned with quantum-native settings; evidence for whether a PQC component provides real value on a realistic computer vision task, evaluated under a standard protocol with a controlled classical baseline, is comparatively scarce.

We focus on a deliberately narrow question: \emph{if a PQC is inserted into an already-competitive classical correlation module, does it change segmentation accuracy -- and if so, is that change attributable to quantum computation specifically?} The individual components that are used are standard: dense support-query correlation is a well-established classical design~\cite{Zhang2021PFENet,Min2021HSNet}, and amplitude-embedded PQCs with a gated residual connection are a straightforward combination of existing quantum-circuit and classical-gating patterns. The main contribution is therefore experimental rather than architectural: a controlled evaluation in which the classical architecture, training protocol, and hyperparameters are held fixed while the presence and type of the refinement module are varied. Critically, the study includes a \emph{parameter-matched classical control} rather than only an unaugmented baseline, allowing any observed effect to be distinguished, in principle, from the generic benefit of adding an extra learned perturbation at that point in the network. We do not claim state-of-the-art few-shot segmentation performance; our classical  proposed baseline is termed as QOSS Corr (Section~\ref{sec:results}) which sits below published methods such as PFENet and HSNet, for reasons we discuss in Section~\ref{sec:limitations}. The QOSS Corr is a quantum-enhanced
segmentation framework centered on support--query correlation refinement. 
QOSS-Corr combines a classical visual feature-extraction pipeline with a
Quantum Correlation Refiner (QCR) that operates on dense correlation
representations.

The main contributions are:
\begin{enumerate}
    \item \textbf{Quantum-engineering characterization.} We quantify the parameter and training-time cost of the quantum module, examine the effect of increasing the measurement dimension, and validate a trained QCR circuit on IBM quantum hardware. The hardware experiment is explicitly treated as a circuit-fidelity check rather than evidence of task-level quantum advantage.
    \item \textbf{A controlled evaluation protocol for quantum-specific effects.} We compare the same classical segmentation pipeline with and without QCR across all four PASCAL-5$^i$ folds and three seeds per fold, while also including a parameter-matched classical MLP refiner and an expanded-measurement quantum variant.
    \item \textbf{A rigorously verified empirical result.} Across 12 matched fold--seed pairs, QCR produces no statistically significant improvement over the classical baseline. The same conclusion is obtained for the parameter-matched MLP control, indicating that the small observed differences are not attributable to the quantum computation under the evaluated configuration.
\end{enumerate}

\section{Related Work}
\label{sec:related}

Few-shot segmentation (FSS) asks a model to segment a novel object class in a query image given only one or a handful of annotated support examples, with no test-time fine-tuning. The problem was formalized by OSLSM~\cite{Shaban2017OSLSM}, whose two-branch conditioning network established the episodic meta-learning protocol -- training and evaluating on disjoint class folds -- that essentially all subsequent FSS work, including PASCAL-5$^i$, still follows.We begin by addressing the quantum engineering domain, which constitutes the core contribution of this paper, before turning to the classical FSS literature that supplies the backbone and baseline this paper's quantum module is inserted into and compared against. Among the FSS methods, prototype-based and dense-correlation methods are the two well-investigated families.

\subsection{Quantum machine learning for vision}PQCs have been proposed as classifier heads, feature encoders, and kernel functions in hybrid classical-quantum pipelines, with variational circuits acting as trainable, differentiable layers dropped into an otherwise classical model~\cite{Biamonte2017QML}. Amplitude embedding and angle embedding are the two most common data-loading strategies for encoding classical feature vectors into qubit states: angle embedding maps each scalar to a single-qubit rotation, so an $n$-qubit circuit loads $n$ values, whereas amplitude embedding loads a normalized $2^n$-dimensional vector into the amplitudes of an $n$-qubit state, at the cost of a harder state-preparation step~\cite{Schuld2021}. Quantum kernel methods, in which a circuit's state overlap serves as a learned similarity function~\cite{Havlicek2019QuantumKernels}, are conceptually the closest prior work to a quantum correlation module, though most published quantum kernel results are on tabular or small image-patch classification data rather than dense per-pixel correlation embedded in a segmentation pipeline.To our knowledge, no prior published work combines a PQC-based correlation-refinement stage with a ResNet-backed few-shot segmenter evaluated on the standard PASCAL-$5^i$ benchmark. Our controlled comparison addresses this specific gap, rather than making a broader claim that quantum methods are unexplored in computer vision---where classification applications, in particular, are relatively mature.

\subsection{Trainability of PQCs} Barren plateaus and other vanishing-gradient phenomena in variational circuits are well documented and are a plausible explanation for weak or flat learning signal in a PQC component embedded in a larger classical model~\cite{McClean2018BarrenPlateaus,Cerezo2021}. Separately, the expressibility and entangling capability of a given ansatz -- how uniformly it can explore the accessible Hilbert space for a fixed depth and qubit count -- has been proposed as a circuit-level descriptor that can help explain, independent of trainability, why a shallow ansatz may or may not carry enough representational capacity for a given task~\cite{Sim2019Expressibility}; we return to both considerations in Section~\ref{sec:discussion}.

\subsection{Prototype-based methods} represent the support set as one or more pooled feature vectors and compare them against query pixels. SG-One~\cite{Zhang2018SGOne} uses masked average pooling to obtain a single support prototype with cosine-similarity guidance. PANet~\cite{Wang2019PANet} introduces prototype alignment regularization: the prototype used to segment the query is also used, symmetrically, to re-segment the support image, enforcing embedding consistency between the two directions. CANet~\cite{Zhang2019CANet} moves beyond a single global prototype with a dense comparison module and an iterative optimization module, adding spatial awareness that pure prototype pooling lacks. A recurring limitation of this family, which motivates the next one, is that collapsing a support object into one or a few vectors discards its spatial structure.

\subsection{Dense correlation methods} instead compute pixel-to-pixel or hyper-correlation matches between support and query feature maps at multiple backbone scales, typically improving accuracy at the cost of a heavier matching module~\cite{Min2021HSNet,Zhang2021PFENet}. HSNet~\cite{Min2021HSNet} builds hypercorrelations across several backbone layers and squeezes the resulting 4D volume with a dedicated convolutional encoder-decoder, and remains one of the strongest non-transformer FSS baselines. DCAMA~\cite{Shi2022DCAMA} and VAT~\cite{Hong2022VAT} replace the 4D-convolutional decoder with, respectively, dense cross-attention and a transformer applied directly to the correlation volume, reflecting the field's broader shift toward attention-based matching. Our classical backbone follows this dense-correlation family, using a simplified two-to-three-scale masked cosine-similarity correlation without full hypercorrelation or 4D convolution -- a deliberate simplification discussed in Section~\ref{sec:limitations}, since it is one reason our classical baseline sits below published HSNet/PFENet numbers.

\section{Method}
\label{sec:method}

\begin{figure*}[!t]
    \centering
    \includegraphics[width=\textwidth]{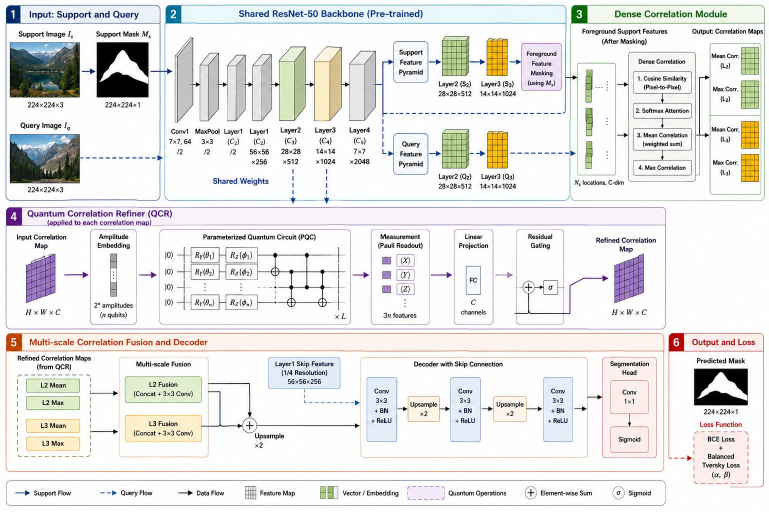}
    \caption{QOSS-Corr architecture illustrating support--query feature extraction, multi-scale dense correlation, quantum correlation refinement, feature fusion, and final mask decoding.}
    \label{fig:architecture}
\end{figure*}

Figure~\ref{fig:architecture} gives a schematic overview of the full QOSS-Corr pipeline described in this section, tracing the support/query inputs through the shared backbone, dense correlation, the quantum refinement module (Section~\ref{sec:method-quantum}), and the decoder, with tensor shapes and the residual/skip paths labeled explicitly. In stages \textbf{(1--2)}, the support and query images (and the support mask) pass through a shared, frozen ResNet-50 backbone; multi-scale feature taps are drawn at Layer-2 ($28{\times}28{\times}512$) and Layer-3 ($14{\times}14{\times}1024$), and the support features are masked by $M_s$ to keep only foreground content, while Layer-1 ($56{\times}56{\times}256$) is retained separately as a decoder skip connection. In stage \textbf{(3)}, the Dense Correlation Module computes a pixel-to-pixel cosine similarity between query and masked support features at each scale, applies softmax attention, and reduces it to mean- and max-correlation maps at both L2 and L3. In stage \textbf{(4)}, each correlation map is passed through the Quantum Correlation Refiner (purple): amplitude embedding maps its $C$-dimensional feature vector at every spatial location onto an $n$-qubit state ($2^n$ amplitudes), two variational layers apply $R_Y$/$R_Z$ rotations with ring-CNOT entanglement, a Pauli readout measures $3n$ features per qubit, and a linear projection maps this back to $C$ channels; a learned scalar residual gate $\sigma$ then combines this correction with the untouched correlation map, $F_c' = F_c + \sigma\,\Delta$ -- this is the only path replaced by an identity map in the classical (no-refiner) condition. In stage \textbf{(5)}, the refined L2 and L3 maps are each fused (concat + $3{\times}3$ conv), summed and upsampled, combined with the Layer-1 skip connection, and decoded through three conv+BN+ReLU / upsample-$\times$2 stages. Finally, in stage \textbf{(6)}, a $1{\times}1$ convolution and sigmoid produce the final $224{\times}224$ predicted mask, trained with the BCE + Balanced Tversky loss of Eqs.~\eqref{eq:bce-loss}--\eqref{eq:total-loss}.

\subsection{Backbone and Multi-Scale Correlation}
\label{sec:method-backbone}

QOSS-Corr uses a ResNet-50 backbone~\cite{He2016ResNet} pretrained on ImageNet~\cite{Deng2009ImageNet}. Features are tapped at three stages: \texttt{layer1} (stride 4, 56$\times$56) is projected to a 32-channel skip connection used only for boundary detail in the decoder; \texttt{layer2} (stride 8, 28$\times$28) and \texttt{layer3} (stride 16, 14$\times$14) are each projected to a shared embedding width $C{=}128$ and used for dense correlation. During training, the backbone is frozen for an initial warm-up period and subsequently fine-tuned at a reduced learning rate (Section~\ref{sec:setup}).

For a pair of support and query feature maps $\mathbf{s}, \mathbf{q} \in \mathbb{R}^{C \times H \times W}$ at a given scale, and a support foreground mask $M \in \{0,1\}^{H\times W}$ resized to that scale, we compute a masked cosine-similarity correlation. Let $\mathcal{F} = \{(i',j') : M_{i',j'} = 1\}$ denote the set of foreground support locations. For each query location $(i,j)$ and each support location $(i',j') \in \mathcal{F}$, the cosine similarity is
\begin{equation}
    \text{sim}\big(\mathbf{q}_{i,j}, \mathbf{s}_{i',j'}\big) = \frac{\mathbf{q}_{i,j} \cdot \mathbf{s}_{i',j'}}{\lVert \mathbf{q}_{i,j} \rVert \, \lVert \mathbf{s}_{i',j'} \rVert}.
    \label{eq:cosine-sim}
\end{equation}
Query locations attend only to foreground support locations, via a softmax over cosine similarity with temperature $\tau$, following the scaled dot-product attention formulation of~\cite{Vaswani2017Attention}:
\begin{equation}
    a_{i,j}(i',j') = \frac{\exp\!\big(\text{sim}(\mathbf{q}_{i,j}, \mathbf{s}_{i',j'})/\tau\big)}{\sum_{(i'',j'') \in \mathcal{F}} \exp\!\big(\text{sim}(\mathbf{q}_{i,j}, \mathbf{s}_{i'',j''})/\tau\big)}
    \label{eq:softmax-attn}
\end{equation}
Where $(i',j') \in \mathcal{F}$, producing an attended support descriptor at each query location by aggregating support features under this attention distribution:
\begin{equation}
    \hat{\mathbf{s}}_{i,j} = \sum_{(i',j') \in \mathcal{F}} a_{i,j}(i',j') \, \mathbf{s}_{i',j'}.
    \label{eq:attended-support}
\end{equation}
We additionally compute two masked similarity statistics at each query location, the mean and max cosine similarity to foreground support locations,
\begin{equation}
\begin{split}
    m_{i,j} = \frac{1}{|\mathcal{F}|}\sum_{(i',j') \in \mathcal{F}} \text{sim}(\mathbf{q}_{i,j}, \mathbf{s}_{i',j'}),\\
    x_{i,j} = \max_{(i',j') \in \mathcal{F}} \text{sim}(\mathbf{q}_{i,j}, \mathbf{s}_{i',j'}).
    \label{eq:mean-max-sim}
\end{split}
\end{equation}
The raw query feature $\mathbf{q}_{i,j}$, the attended support descriptor $\hat{\mathbf{s}}_{i,j}$, and the two scalar statistics $m_{i,j}, x_{i,j}$ are concatenated channel-wise and fused by a $3{\times}3$ convolution with weights $W_{\text{fuse}}$ into an out-of-scale correlation map $\mathbf{c} \in \mathbb{R}^{C'\times H \times W}$, $C'{=}64$:
\begin{equation}
    \mathbf{c}_{i,j} = W_{\text{fuse}} * \big[\mathbf{q}_{i,j} \, \Vert \, \hat{\mathbf{s}}_{i,j} \, \Vert \, m_{i,j} \, \Vert \, x_{i,j}\big].
    \label{eq:correlation-fusion}
\end{equation}
Including the raw query feature in this fusion -- rather than only the support-attended descriptor -- ensures the decoder has direct access to the query image's own appearance, not solely a support-reconstructed version of it.

\subsection{Quantum Correlation Refiner}
\label{sec:method-quantum}

The Quantum Correlation Refiner (QCR) is applied, independently, to the correlation map at each backbone scale. Given a correlation map $\mathbf{c} \in \mathbb{R}^{C'\times H \times W}$ with $C' = 2^{n}$ for $n$ qubits ($C'{=}64$, $n{=}6$ in our experiments), each spatial location's $C'$-dimensional channel vector is treated as one classical sample and amplitude-embedded~\cite{Schuld2021} into an $n$-qubit state:
\begin{equation}
    |\psi(\mathbf{x})\rangle = \text{AmplitudeEmbed}(\mathbf{x}), \quad \mathbf{x} \in \mathbb{R}^{C'}, \ \|\mathbf{x}\|=1 \text{ (norm)}
\end{equation}
A trainable, hardware-efficient ansatz~\cite{Kandala2017VQE} of $L{=}2$ entangling layers is applied, each consisting of per-qubit $R_Y$ and $R_Z$ rotations followed by a ring of CNOT gates. The circuit unitary is given by:
\begin{equation}
    U(\boldsymbol{\theta}) = \prod_{l=1}^{L} \left( \prod_{q} \text{CNOT}_{q,(q+1) \bmod n} \right) \left( \prod_q R_Z(\theta_{l,q,2}) R_Y(\theta_{l,q,1}) \right)
\end{equation}
Here, $\boldsymbol{\theta}$ denotes the full set of trainable circuit
parameters, $L$ is the number of variational layers, $n$ is the number
of qubits, and $q \in \{0, \dots, n-1\}$ indexes the qubits; within
each layer $l$, $R_Y(\theta_{l,q,1})$ and $R_Z(\theta_{l,q,2})$ apply
single-qubit rotations parameterized by angles $\theta_{l,q,1}$ and
$\theta_{l,q,2}$, respectively, followed by a ring of
$\text{CNOT}_{q,(q+1) \bmod n}$ gates that entangle each qubit with its
cyclic neighbor. After applying $U(\boldsymbol{\theta})$, the circuit is measured
via Pauli-Z expectation values, $\langle Z_0 \rangle, \dots, \langle Z_{n-1}\rangle \in [-1,1]^n$, read out in the manner of a circuit-centric quantum classifier~\cite{SchuldBocharov2020CircuitCentric} and linearly projected back to $C'$ channels and added as a \emph{gated residual}:
\begin{equation}
    \mathbf{c}'_{i,j} = \mathbf{c}_{i,j} + \sigma_{\text{res}} \cdot W_{\text{out}}\big(\langle Z \rangle(\mathbf{c}_{i,j}; \boldsymbol\theta)\big),
\end{equation}
where $\sigma_{\text{res}}$ is a single learned scalar, initialized to $0.1$, and $(i,j)$ indexes spatial location. The small initialization ensures the classical correlation signal dominates early in training, with the model free to increase $\sigma_{\text{res}}$ if the quantum correction proves useful. Circuit simulation uses PennyLane's~\cite{Bergholm2018PennyLane} \texttt{default.qubit} device with the \texttt{backprop} differentiation method, which vectorizes the per-pixel circuit evaluation across the batch and spatial dimensions; we found this substantially faster than \texttt{lightning}-based adjoint differentiation for the small qubit counts and per-pixel batch sizes used here.

We apply QCR to the two coarser correlation scales (\texttt{layer2}, \texttt{layer3}) and not the finest scale, to bound the number of per-batch circuit evaluations to a computationally tractable range.

\subsection{Proposed extension: expanded Pauli measurement (QCR-XYZ)}
\label{sec:qcr-xyz}
One way to ease the measurement-dimensionality bottleneck described
above without running any extra circuits is to read out all
three single-qubit Pauli expectation values per qubit, instead of
just $\langle Z \rangle$. This relates to the circuit-expressibility
question raised in~\cite{SchuldBocharov2020CircuitCentric}. Concretely, we measure
\begin{equation}
\begin{split}
\langle P \rangle(c_{i,j}; \boldsymbol{\theta}) =
    \big[\langle X_0 \rangle, \dots, \langle X_{n-1} \rangle,\;
    \langle Y_0 \rangle, \dots, \langle Y_{n-1} \rangle,\;
    \langle Z_0 \rangle, \\ \dots, \langle Z_{n-1} \rangle\big]
    \in [-1, 1]^{3n}
\end{split}
\end{equation}
Here, $n$ denotes the number of qubits, $X_q$, $Y_q$, and $Z_q$ are
the Pauli operators measured on qubit $q \in \{0, \dots, n-1\}$,
$\boldsymbol{\theta}$ are the trainable circuit parameters, and
$c_{i,j}$ is the correlation feature at spatial location $(i,j)$
used to prepare the input quantum state; $\langle P \rangle(c_{i,j};
\boldsymbol{\theta})$ is the resulting $3n$-dimensional vector of
Pauli expectation values, each bounded in $[-1, 1]$.

\label{sec:methodology-remark-on-bottleneck}
This triples the number of measured quantities per spatial location,
from $n{=}6$ to $3n{=}18$. This is still much smaller than the
64-dimensional correlation space, and far smaller than the $4^n - 1 =
4095$ measurements that full state tomography would require for
$n{=}6$ qubits. In other words, this change raises the ceiling on
expressiveness noted earlier, but does not remove it entirely.

Because all three expectation values come from the same state vector
produced by a single circuit run under \texttt{default.qubit} with
backprop simulation, this adds no extra circuit evaluations in
simulation. On real hardware or shot-based execution, however, each
measurement basis requires its own circuit run, so this trade-off
would need to be reassessed in that setting.

The readout projection and gated residual, otherwise identical in form to the single-observable case above, become
\begin{equation}
    W_{\text{out}} \in \mathbb{R}^{64 \times 3n}, \qquad
    \mathbf{c}'_{i,j} = \mathbf{c}_{i,j} + \sigma_{\text{res}} \cdot W_{\text{out}}\big(\langle P \rangle(\mathbf{c}_{i,j}; \boldsymbol\theta)\big),
    \label{eq:xyz-residual}
\end{equation}
with every other component of QCR (ansatz, layer count, embedding, gate initialization) unchanged from Section~\ref{sec:method-quantum}. This adds $64 \times (3n - n) = 768$ trainable parameters to $W_{\text{out}}$ (for $n{=}6$) relative to the $Z$-only variant, still negligible next to the backbone (Section~\ref{sec:complexity}). 
\subsection{Decoder}
\label{sec:method-decoder}

Correlation maps from all active scales are bilinearly upsampled to the coarsest-common resolution, concatenated, and fused by a $3{\times}3$ convolution. The result is progressively upsampled to full input resolution via three transposed-convolution stages, with the \texttt{layer1} skip connection concatenated in after the first upsampling stage to recover boundary detail. A final $1{\times}1$ convolution produces the segmentation logit map; its bias is initialized to the logit of the dataset's empirical foreground prior ($\approx 0.10$--$0.30$ depending on fold) to accelerate early convergence.

\subsection{Algorithm Summary}
\label{sec:method-algorithm}

The complete per-episode forward pass (backbone $\to$ correlation $\to$ QCR $\to$ decoder $\to$ loss) and the epoch-level training loop are given as detailed pseudocode in Appendix~\ref{app:algorithms} (Algorithms~\ref{alg:qoss-corr} and~\ref{alg:training-loop}), rather than in the main text, to keep this section focused on the architectural description above. In summary: each episode's support and query images pass through the shared frozen backbone; correlation and QCR are applied per scale as described in Sections~\ref{sec:method-backbone}--\ref{sec:method-quantum}; the result is fused and decoded to a segmentation mask; and the loss is BCE + Tversky (Eqs.~\eqref{eq:bce-loss}--\eqref{eq:total-loss}). In the classical (no-refiner) condition, the QCR step is replaced by the identity map, matching the red identity path in Figure~\ref{fig:architecture}.

\section{Experimental Setup}
\label{sec:setup}

\subsection{Dataset and protocol} We evaluate on PASCAL-5$^i$, the standard 1-shot few-shot segmentation benchmark derived from PASCAL VOC 2012~\cite{Everingham2010PascalVOC}. The 20 object classes are partitioned into four folds of five classes each (Fold 0: aeroplane, bicycle, bird, boat, bottle; Fold 1: bus, car, cat, chair, cow; Fold 2: diningtable, dog, horse, motorbike, person; Fold 3: pottedplant, sheep, sofa, train, tvmonitor), following the standard split used throughout the few-shot segmentation literature~\cite{Shaban2017OSLSM}; for each fold, the model is trained on episodes drawn from the other three folds' classes (15 classes) and evaluated on episodes drawn from the held-out fold's five classes, which the model never sees during training. We report class-mean IoU (mIoU), computed by accumulating per-class intersection and union across the full validation episode set before averaging across classes, following standard practice.

\subsection{Training details} Images are resized to $224{\times}224$. We train with AdamW~\cite{Loshchilov2019AdamW}, base learning rate $3{\times}10^{-4}$ for the classical parameters, $9{\times}10^{-6}$ for the backbone once unfrozen, and (in quantum-enabled runs) $3{\times}10^{-4}$ for the PQC parameters, in a separate optimizer parameter group. The loss is a sum of binary cross-entropy and a Tversky loss~\cite{Salehi2017Tversky} (\(\alpha{=}\beta{=}0.5\), i.e. balanced Dice) unless otherwise noted. Writing $\hat{y}_{i,j} \in [0,1]$ for the predicted foreground probability at pixel $(i,j)$ and $y_{i,j} \in \{0,1\}$ for the ground-truth label, the binary cross-entropy term over the $N{=}H{\times}W$ pixels of the query mask is:
\begin{equation}
    \mathcal{L}_{\text{BCE}} = -\frac{1}{N} \sum_{i,j} \Big[ y_{i,j} \log \hat{y}_{i,j} + (1 - y_{i,j}) \log (1 - \hat{y}_{i,j}) \Big]
    \label{eq:bce-loss}
\end{equation}
The Tversky index~\cite{Salehi2017Tversky} generalizes the Dice coefficient by separately weighting false negatives and false positives via $\alpha, \beta \ge 0$:
\begin{equation}
\begin{split} 
    \text{TI} = \frac{\sum_{i,j} \hat{y}_{i,j}\, y_{i,j}}{\sum_{i,j} \hat{y}_{i,j}\, y_{i,j} + \alpha \sum_{i,j} (1-\hat{y}_{i,j})\, y_{i,j} + \beta \sum_{i,j} \hat{y}_{i,j}\, (1-y_{i,j})} \,,
    \\
 \mathcal{L}_{\text{Tversky}} = 1 - \text{TI},
    \label{eq:tversky-loss}
\end{split}
\end{equation}
which reduces to the standard (balanced) Dice loss at $\alpha{=}\beta{=}0.5$, as used in all runs reported here. The total training objective is
\begin{equation}
    \mathcal{L} = \mathcal{L}_{\text{BCE}} + \mathcal{L}_{\text{Tversky}}.
    \label{eq:total-loss}
\end{equation} The backbone is frozen for the first 15 epochs; all runs reported here train for 15 total epochs with early stopping (patience 10), so the backbone remains frozen for the entirety of every reported run. Batch size is 16 episodes; support and query images are augmented independently with random scale jitter, horizontal flip, and color jitter during training. Validation uses a fixed, seeded set of 1000 episodes per fold (unaugmented, flip test-time-augmented) to ensure classical and quantum runs on the same fold are evaluated on identical data.
\subsection{Reported runs} We report 3 random seeds (42, 43, 44) per condition per fold, for both the classical (no-refiner) and quantum-augmented conditions across all four PASCAL-5$^i$ folds (12 pairs total), and additionally for the parameter-matched MLP-refiner and expanded-measurement QCR-XYZ conditions on Fold 1 (Section~\ref{sec:ablation}). All results in this paper are verified directly against the original training logs (Section~\ref{sec:results}).

The epoch-level training procedure common to every run reported in this paper is given in Appendix~\ref{app:algorithms}, Algorithm~\ref{alg:training-loop}, building on the single-episode forward pass of Algorithm~\ref{alg:qoss-corr}. 

\section{Results}
\label{sec:results}

\subsection{4-Fold Comparison, 3 Seeds per Condition}

Table~\ref{tab:main} reports validation mIoU, mean $\pm$ std over 3 seeds (42, 43, 44), verified against original training logs, for the three directly-comparable conditions -- classical (no refiner), quantum QCR ($Z$-only), and the parameter-matched classical MLP-refiner control -- on all four folds; Table~\ref{tab:per-seed} gives the individual seed values underlying it. The expanded-measurement quantum variant QCR-XYZ, verified on all four folds, is reported separately in Table~\ref{tab:xyz-allfolds} (Section~\ref{sec:ablation}). A parameter-free noise-injection control is verified on all four folds (Table~\ref{tab:noise}).

Pooled across all 4 folds $\times$ 3 seeds $=12$ matched pairs, QCR produces a near-zero mean difference relative to the classical baseline: $\Delta = +0.0001$ mIoU, $t(11) = 0.10$, $p = 0.92$, Cohen's $d_z = 0.03$~\cite{Cohen1988}. The four per-fold mean differences straddle zero ($-0.0007$, $-0.0030$, $-0.0010$, and $+0.0052$ for Folds 0--3, respectively), with no consistent direction across folds. Because the paired test does not reject the null hypothesis, we do not interpret this result as evidence of improvement or deterioration. The reported confidence interval is presented as uncertainty around the estimated difference, not as evidence of equivalence. The pooled comparison is therefore treated as the primary empirical result.

\subsubsection{Per-fold heterogeneity in the quantum QCR deltas} These four per-fold mean deltas straddle zero with no discernible pattern tied to fold difficulty: Fold 0 ($-0.0007$) and Fold 2 ($-0.0010$) are both essentially flat; Fold 1 ($-0.0030$) is the only fold with a consistent within-fold sign, and it is negative; Fold 3 ($+0.0052$) is the only fold with a positive mean delta, and it is also the fold with the lowest absolute mIoU of the four (Table~\ref{tab:main}) and the largest per-seed spread in its own deltas ($+0.0031, +0.0023, +0.0101$) -- one seed (44) accounts for most of that fold's positive mean. None of the four individual per-fold comparisons reaches significance at $n=3$, and we have no reason to treat any one fold's direction as more informative than another's: the pooled result is the one number in this paper we would point to first. The MLP-refiner and QCR-XYZ controls (Table~\ref{tab:main}, Table~\ref{tab:xyz-allfolds}) show broadly the same per-fold pattern reproduced by controls with no PQC or with a wider PQC readout, which bears directly on whether these small per-fold deltas are quantum-specific.

The parameter-matched MLP-refiner control, run on all four folds under the identical protocol, not just Fold 1 leads to the same conclusion: pooled across the same 12 pairs, mean $\Delta$ (MLP-refiner $-$ classical) $= +0.0003$ mIoU, $t(11) = 0.09$, $p = 0.93$, $d_z = 0.03$, and quantum vs.\ MLP-refiner directly is likewise null (mean $\Delta = -0.0002$, $t(11) = -0.07$, $p = 0.95$). Beyond the pooled tests, the per-fold MLP-refiner deltas track quantum's own per-fold deltas in sign on all four folds (Table~\ref{tab:main}: both roughly flat on Fold 0, both negative on Folds 1--2, both most positive on Fold 3, the fold with the lowest absolute mIoU). If the quantum computation contributed an effect beyond the generic MLP perturbation, we would expect the two refiners' per-fold deltas to decouple in sign on at least some folds; instead they co-vary with fold identity, consistent with fold-level idiosyncrasies in the gated-residual training dynamics driving the small deltas observed, rather than any property of the correction itself, quantum or classical.

\begin{table*}[!t]
\centering
\caption{Validation mIoU by fold and condition, mean $\pm$ std over 3 seeds (42, 43, 44). $\Delta_{Q}$ is quantum QCR minus classical; $\Delta_{M}$ is MLP-refiner minus classical}
\label{tab:main}
\begin{tabular}{lccccc}
\toprule
Fold (held-out classes) & Classical & + QCR & + MLP-refiner & $\Delta_{Q}$ & $\Delta_{M}$ \\
\midrule
Fold 0 (1--5)   & 0.4637 $\pm$ 0.0072 & 0.4630 $\pm$ 0.0035 & 0.4656 $\pm$ 0.0062 & $-0.0007$ & $+0.0018$ \\
Fold 1 (6--10)  & 0.6053 $\pm$ 0.0125 & 0.6024 $\pm$ 0.0127 & 0.6027 $\pm$ 0.0125 & $-0.0030$ & $-0.0026$ \\
Fold 2 (11--15) & 0.4729 $\pm$ 0.0156 & 0.4719 $\pm$ 0.0131 & 0.4687 $\pm$ 0.0098 & $-0.0010$ & $-0.0042$ \\
Fold 3 (16--20) & 0.4143 $\pm$ 0.0144 & 0.4195 $\pm$ 0.0133 & 0.4205 $\pm$ 0.0116 & $+0.0052$ & $+0.0061$ \\
\midrule
\textbf{Pooled (12 pairs)} & \textbf{0.4891} & \textbf{0.4892} & \textbf{0.4894} & \textbf{+0.0001} & \textbf{+0.0003} \\
\bottomrule
\end{tabular}
\\[0.3em]

\end{table*}

\subsection{The expanded-measurement variant, QCR-XYZ} Table~\ref{tab:xyz-allfolds} adds this variant to the three conditions in Table~\ref{tab:main}. Pooled across the same 12 fold$\times$seed pairs, QCR-XYZ is statistically indistinguishable from the no-refiner baseline: mean $\Delta = +0.0005$ mIoU, $t(11) = 0.24$, $p = 0.81$, $d_z = 0.07$ -- a null result of the same character as $Z$-only quantum's own pooled null. On Fold 1 specifically, all four conditions fall within a 0.0029 mIoU band of each other, well inside one pooled standard deviation. Per-seed differences, which are more informative than the means alone at $n=3$ (Table~\ref{tab:per-seed}), tell a more specific story there: $Z$-only quantum trails the no-refiner baseline by a small, consistent, low-variance margin on every one of the three seeds ($-0.0029$, $-0.0032$, $-0.0028$; range $0.0004$), while the MLP-refiner's and QCR-XYZ's differences from the baseline flip sign across seeds and show 15--20$\times$ more spread ($mlp$: $+0.0015$, $-0.0021$, $-0.0072$; $XYZ$: $-0.0016$, $+0.0002$, $-0.0047$), consistent with noise rather than a real effect in either direction. Head-to-head, quantum vs.\ MLP-refiner ($-0.0044$, $-0.0011$, $+0.0044$) and quantum vs.\ QCR-XYZ ($+0.0013$, $+0.0034$, $-0.0019$) both flip sign across seeds and center near zero, no consistent ordering among the three refiner conditions on Fold 1. Tripling the measurement degrees of freedom from 6 to 18 (Section~\ref{sec:qcr-xyz}) did not reproduce $Z$-only quantum's small, consistent negative effect, nor did it produce a positive one, on Fold 1 or pooled; this weakens, without eliminating, the measurement-dimensionality bottleneck (Section~\ref{sec:methodology-remark-on-bottleneck}) as a complete explanation for the module's lack of benefit, since relaxing it did not measurably change the outcome on any fold. Fold 1 is the one fold in Table~\ref{tab:main} that shows a consistent-sign effect at all for $Z$-only quantum (trailing classical on all 3 seeds) but at $-0.0030$ mean, it is also the smallest-magnitude of the four per-fold deltas, and this sign does not generalize pooled.

\begin{table*}[!t]
\centering
\caption{Quantum QCR-XYZ (expanded measurement), all four folds, mean $\pm$ std over 3 seeds (42, 43, 44). $\Delta$ is QCR-XYZ minus classical.}
\label{tab:xyz-allfolds}
\begin{tabular}{lccc}
\toprule
Fold (held-out classes) & Classical & + QCR-XYZ & $\Delta$ \\
\midrule
Fold 0 (1--5)   & 0.4637 $\pm$ 0.0072 & 0.4650 $\pm$ 0.0054 & $+0.0012$ \\
Fold 1 (6--10)  & 0.6053 $\pm$ 0.0125 & 0.6033 $\pm$ 0.0106 & $-0.0020$ \\
Fold 2 (11--15) & 0.4729 $\pm$ 0.0156 & 0.4689 $\pm$ 0.0119 & $-0.0040$ \\
Fold 3 (16--20) & 0.4143 $\pm$ 0.0144 & 0.4211 $\pm$ 0.0126 & $+0.0068$ \\
\midrule
\textbf{Pooled (12 pairs)} & \textbf{0.4891} & \textbf{0.4896} & \textbf{+0.0005} \\
\bottomrule
\end{tabular}
\\[0.3em]
\end{table*}

\subsubsection{A fixed-noise, parameter-free control} 
\begin{figure*}[!t]
    \centering
    \includegraphics[width=\textwidth]{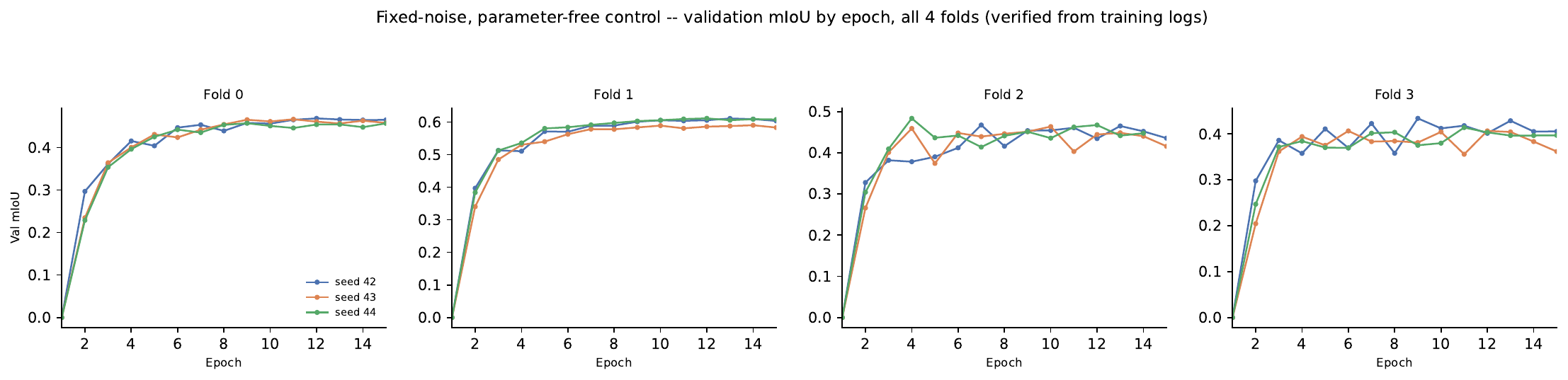}
    \caption{Validation mIoU by epoch for the fixed-noise, parameter-free control on all four folds (3 seeds each), extracted directly from the underlying training logs.}
    \label{fig:noise-folds01}
\end{figure*}

This condition tests whether \emph{any} perturbation, structured or not, is gated similarly at the same residual location; it is verified on all four folds directly from per-epoch training logs (Table~\ref{tab:noise}, Figure~\ref{fig:noise-folds01}), superseding the earlier state in which Folds 2 and 3 rested on final-value summaries alone. Pooled across all 12 fold$\times$seed pairs, noise-injection is indistinguishable from the classical baseline (mean $\Delta = +0.0003$ mIoU, $t(11) = 0.29$, $p = 0.78$, $d_z=0.08$), and on the two folds where quantum QCR is also directly comparable, from quantum QCR itself ($t(5)=0.93$, $p=0.40$, Folds 2--3). Across all four folds now checked, the network does not measurably reward or penalize an unstructured perturbation any differently than it does the quantum or classical structured ones, at the level of final validation mIoU. The gate can learn to suppress a signal it judges structurally uninformative and still land at a final accuracy statistically indistinguishable from the other conditions.

\begin{table*}[!t]
\centering
\caption{Fixed-noise, parameter-free control, verified on all four folds, mean $\pm$ std over 3 seeds (42, 43, 44), $\Delta$ is noise minus classical.}
\label{tab:noise}
\begin{tabular}{lccc}
\toprule
Fold (held-out classes) & Classical & + Noise-injection & $\Delta$ \\
\midrule
Fold 0 (1--5)   & 0.4637 $\pm$ 0.0072 & 0.4639 $\pm$ 0.0061 & $+0.0001$ \\
Fold 1 (6--10)  & 0.6053 $\pm$ 0.0125 & 0.6042 $\pm$ 0.0121 & $-0.0011$ \\
Fold 2 (11--15) & 0.4729 $\pm$ 0.0156 & 0.4714 $\pm$ 0.0106 & $-0.0015$ \\
Fold 3 (16--20) & 0.4143 $\pm$ 0.0144 & 0.4179 $\pm$ 0.0141 & $+0.0036$ \\
\midrule
\textbf{Pooled (12 pairs)} & \textbf{0.4891} & \textbf{0.4893} & \textbf{+0.0003} \\
\bottomrule
\end{tabular}
\\[0.3em]

\end{table*}

\begin{figure*}[!t]
    \centering
    \includegraphics[width=0.7\textwidth]{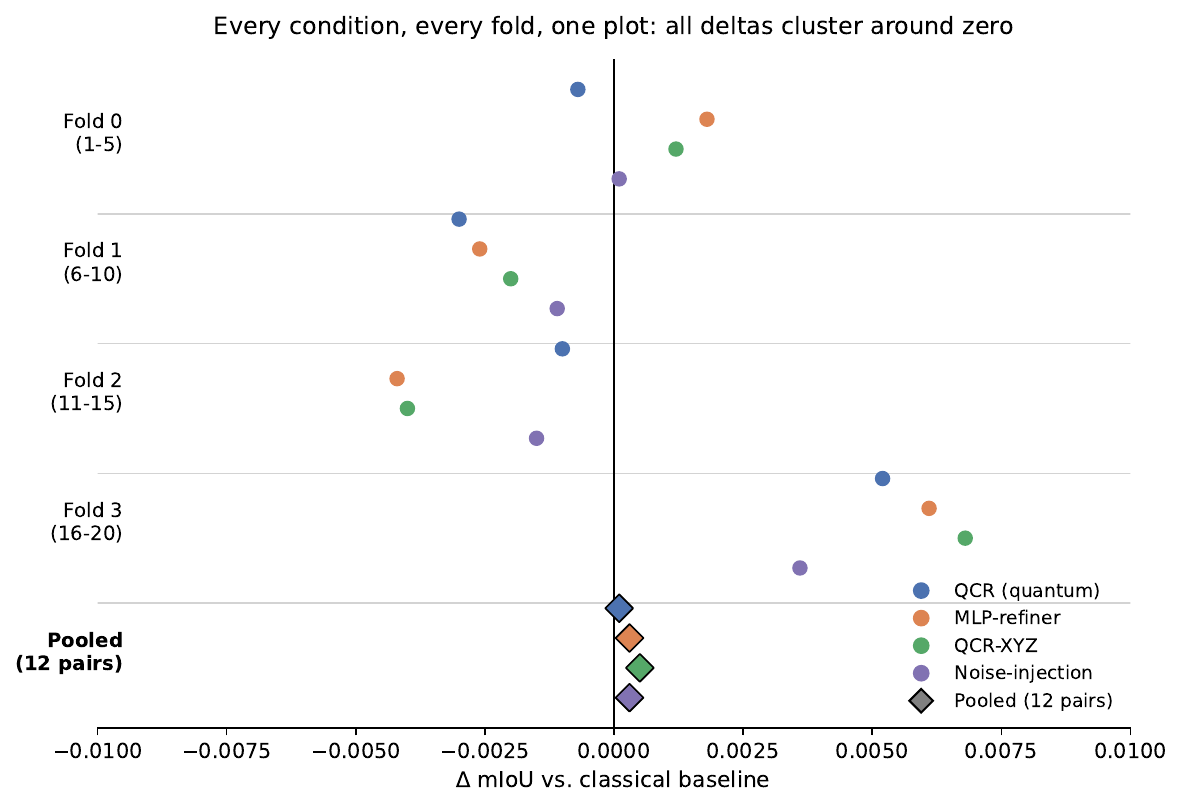}
    \caption{ All condition-vs.-classical deltas ($\Delta$ mIoU), every fold and the pooled estimate, in one plot -- consolidating Tables 1--3. Circles are per-fold point estimates (QCR, MLP-refiner, QCR-XYZ, noise-injection); diamonds are the corresponding pooled (12-pair) estimates.}
    \label{fig:fold-comparison}
\end{figure*}

\begin{figure*}[!t]
    \centering
    \includegraphics[width=\textwidth]{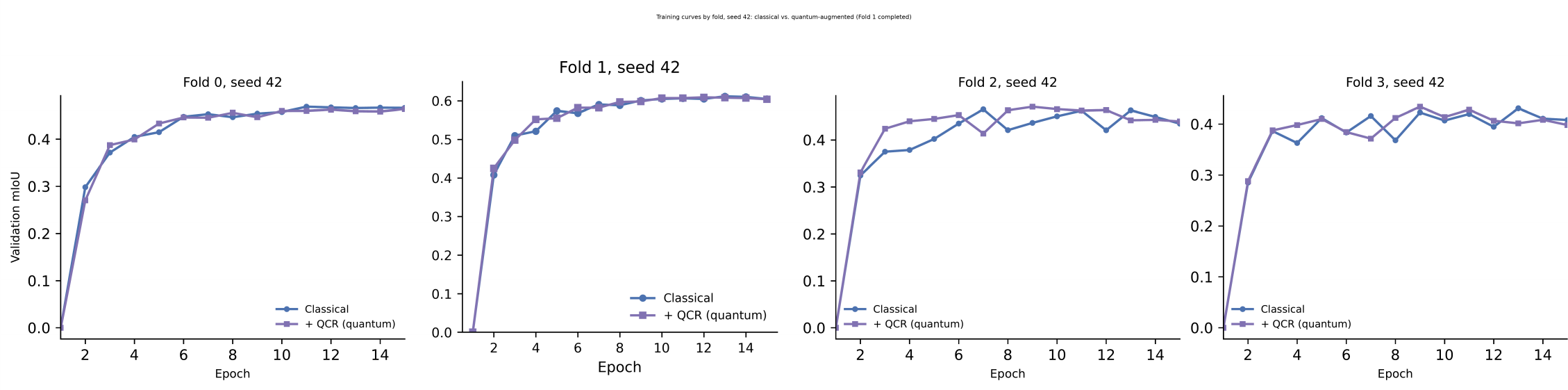}
    \caption{Validation mIoU by epoch, classical vs.\ quantum-augmented (Z-only QCR), seed 42, all four folds, extracted directly from verified training logs. The backbone remains frozen for the full 15-epoch run in every condition shown. }
    \label{fig:fold3-curves}
\end{figure*}

Figure ~\ref{fig:fold-comparison} consolidates every fold-level and pooled comparison from Tables 1–3 into a single view: all four conditions' deltas against the classical baseline, all four folds, plus the pooled result, plotted against a common zero line. Rather than requiring a reader to cross-reference three separate tables to see it, the pattern this section argues for is directly visible here — every point, across every condition and every fold, sits within roughly ±0.007 mIoU of zero, with no condition (quantum or classical) separating itself from the others at any fold, and all four pooled estimates (diamonds) collapsing to within 0.0004 mIoU of exactly zero.
\begin{table*}[!t]
\centering
\caption{Per-seed validation mIoU underlying Table~\ref{tab:main}, Table~\ref{tab:xyz-allfolds}, and Table~\ref{tab:noise}, all four folds, verified against training logs. Noise-injection is verified on all four folds.}
\label{tab:per-seed}
\begin{tabular}{llccccc}
\toprule
Fold & Seed & Classical & + QCR & + MLP-refiner & + QCR-XYZ & + Noise \\
\midrule
\multirow{3}{*}{Fold 0 (1--5)}
  & 42 & 0.4689 & 0.4640 & 0.4726 & 0.4706 & 0.4684 \\
  & 43 & 0.4668 & 0.4659 & 0.4607 & 0.4644 & 0.4663 \\
  & 44 & 0.4555 & 0.4592 & 0.4634 & 0.4599 & 0.4569 \\
\midrule
\multirow{3}{*}{Fold 1 (6--10)}
  & 42 & 0.6116 & 0.6087 & 0.6131 & 0.6100 & 0.6110 \\
  & 43 & 0.5909 & 0.5877 & 0.5888 & 0.5911 & 0.5902 \\
  & 44 & 0.6134 & 0.6106 & 0.6062 & 0.6087 & 0.6113 \\
\midrule
\multirow{3}{*}{Fold 2 (11--15)}
  & 42 & 0.4654 & 0.4709 & 0.4799 & 0.4725 & 0.4673 \\
  & 43 & 0.4625 & 0.4594 & 0.4644 & 0.4556 & 0.4635 \\
  & 44 & 0.4908 & 0.4855 & 0.4619 & 0.4785 & 0.4835 \\
\midrule
\multirow{3}{*}{Fold 3 (16--20)}
  & 42 & 0.4310 & 0.4341 & 0.4338 & 0.4334 & 0.4337 \\
  & 43 & 0.4058 & 0.4081 & 0.4126 & 0.4082 & 0.4064 \\
  & 44 & 0.4062 & 0.4163 & 0.4150 & 0.4218 & 0.4137 \\
\bottomrule
\end{tabular}
\end{table*}
Figure~\ref{fig:fold3-curves} extends this comparison to all four folds' seed-42 training curves, classical vs.\ quantum, regenerated directly from verified per-epoch logs. Folds 0, 1, 2 and 3 show matched classical and quantum curves tracking closely throughout training and converging to within a small margin by the final epochs, consistent with the pooled null result (Table~\ref{tab:main}). 
\subsection{Learned Gate Behavior}
\label{sec:gate-behavior}
\begin{figure*}[!t]
    \centering
    \includegraphics[width=\textwidth]{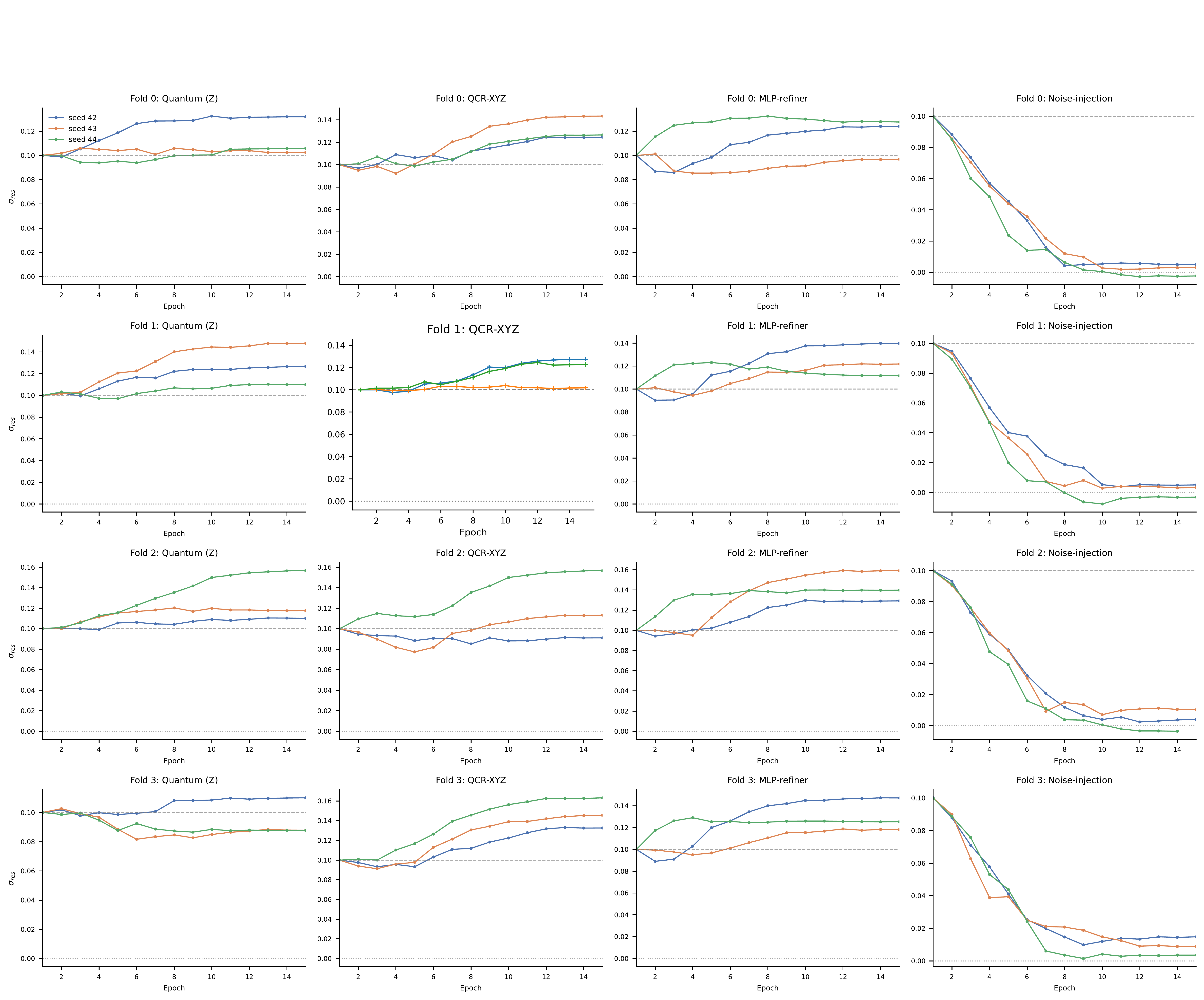}
    \caption{Per-epoch trajectory of the learned residual gate $\sigma_{\text{res}}$, all four folds $\times$ all four refiner conditions (quantum $Z$-only, QCR-XYZ, MLP-refiner, noise-injection), 3 seeds each.}
    \label{fig:residual-scale}
\end{figure*}

Figure~\ref{fig:residual-scale} now plots the complete per-epoch $\sigma_{\text{res}}$ trajectory for all four conditions -- quantum ($Z$-only QCR), QCR-XYZ, MLP-refiner, and noise-injection -- across all four folds (3 seeds each, 48 runs totals). s grid: on every fold, all three structured refiners -- quantum, QCR-XYZ, and MLP-refiner -- drift upward from the $0.1$ initialization into a broadly overlapping band (roughly $0.09$--$0.17$ depending on fold and seed), while noise-injection's gate declines toward or below zero on every fold, with Fold 3 the one partial exception where it ends slightly \emph{above} zero rather than crossing it ($0.004$--$0.015$, mean $0.009$) -- the same fold where noise-injection's mean final mIoU exceeded classical's (Table~\ref{tab:noise}).

On Fold 1, where per-epoch logs are most complete: quantum ends at $0.11$--$0.15$ (mean $0.128$) and MLP at $0.11$--$0.14$ (mean $0.124$), while noise-injection ends at $-0.003$ to $+0.005$ (mean $0.002$) -- a clear separation between structured and unstructured perturbations. On Fold 3, quantum's three seeds split, with one ending above initialization ($0.110$) and two ending below it ($0.088$ each, mean $0.095$); MLP's three seeds all end above initialization ($0.118$--$0.147$, mean $0.130$). It is evident that quantum, QCR-XYZ, and MLP-refiner track each other's upward drift on both folds as well, while noise-injection's gate separates from all three there too.

\subsection{Ablation Study}
\label{sec:ablation}

tHE central open question raised by these results is whether the observed mIoU difference can be attributed to quantum computation specifically, or merely to the presence of an additional gated residual perturbation at that point in the network. This section examines that comparison across four conditions:

\begin{itemize}
    \item \textbf{Classical (no refiner):} The QOSS-Corr baseline with QCR removed entirely (identity passthrough).
    \item \textbf{Classical + QCR (quantum, $Z$-only).} The originally-proposed quantum-augmented model: 6-qubit amplitude embedding, 2 variational layers, single-qubit $\langle Z\rangle$ readout (Section~\ref{sec:method-quantum}).
    \item \textbf{Classical + MLP-refiner (param-matched):} QCR's PQC replaced by a small classical MLP with a parameter count matched to $\boldsymbol\theta$ (the PQC's rotation angles) plus $W_{\text{out}}$, under the identical gated-residual scheme and $\sigma_{\text{res}}$ initialization. Verified across all four folds (Table~\ref{tab:main}, Section~\ref{sec:results}).
    \item \textbf{Classical + QCR-XYZ (expanded measurement):} The variant specified in Section~\ref{sec:qcr-xyz}: identical circuit, but reading $\langle X\rangle,\langle Y\rangle,\langle Z\rangle$ per qubit (18 real degrees of freedom instead of 6), directly testing whether relaxing the measurement-dimensionality bottleneck (Section~\ref{sec:methodology-remark-on-bottleneck}) changes the outcome. Verified across all four folds (Table~\ref{tab:xyz-allfolds}).
\end{itemize}

All four conditions are verified against training logs across all four folds and reported together: the first three in Table~\ref{tab:main} (Section~\ref{sec:results}), QCR-XYZ in Table~\ref{tab:xyz-allfolds}, together with the cost analysis that motivated it. A fixed-noise, parameter-free control is now verified on all four folds (Table~\ref{tab:noise}).

On Fold 1 specifically, all four conditions (Table~\ref{tab:main} plus Table~\ref{tab:xyz-allfolds}) are mutually statistically indistinguishable -- every mean falls within a 0.0029 mIoU band, well inside one pooled standard deviation. Two comparisons are informative beyond the raw means, using per-seed differences rather than means alone (Section~\ref{sec:results}): quantum vs.\ no-refiner shows a small, consistent, low-variance decrease on every one of 3 seeds ($-0.0029, -0.0032, -0.0028$), while quantum vs.\ MLP-refiner---the comparison that would isolate a genuine quantum-specific effect from a generic learned-perturbation effect---flips sign across seeds ($-0.0044, -0.0011, +0.0044$) and shows no consistent ordering, a pattern that also holds pooled across all four folds (mean $\Delta=-0.0002$, $t(11)=-0.07$, $p=0.95$, Table~\ref{tab:main}). Together these indicate the param-matched classical control performs comparably to the quantum refiner on every fold tested, which is not consistent with any effect being attributable to quantum computation specifically rather than to the generic presence of a small trained perturbation at this location. Furthermore, QCR-XYZ vs.\ quantum ($Z$-only) also flips sign across seeds ($+0.0013, +0.0034, -0.0019$), showing that tripling the measurement degrees of freedom did not reproduce $Z$-only quantum's small, consistent negative effect, nor produce a positive one, weakening (without eliminating) the measurement-dimensionality bottleneck as a complete explanation for the module's lack of benefit.

\subsection{Cost note on QCR-XYZ} Reading three Pauli observables per qubit instead of one adds no additional circuit executions in \texttt{default.qubit}+\texttt{backprop} simulation. All three expectation values come from the same state vector (Section~\ref{sec:qcr-xyz}) but this does not mean the variant is free. Measured training time was $\approx$144--149s/epoch for QCR-XYZ versus $\approx$89--93s/epoch for $Z$-only quantum on the same hardware and protocol, a $\approx$60\% increase, plausibly from \texttt{backprop} differentiating through three times as many measurement operators even though the forward circuit itself runs once. This is a real, reproducible cost increase for a result that is statistically indistinguishable from the cheaper $Z$-only variant. Peak GPU memory was also higher for QCR-XYZ in these logs (0.9GB vs.\ 0.7GB), though this figure is not treated as reliable on its own, since VRAM readings across conditions in the same long-running session can be affected by allocator caching and fragmentation independent of the model itself.

\subsection{Computational Complexity}
\label{sec:complexity}

Table~\ref{tab:complexity} reports the parameter cost of each architectural component, computed analytically from the layer dimensions given in Section~\ref{sec:method}. The ResNet-50 backbone (frozen throughout all reported runs, Section~\ref{sec:limitations}) is by far the dominant cost. The Quantum Correlation Refiner itself is parametrically tiny: two variational layers over 6 qubits, each contributing one $R_Y$ and one $R_Z$ rotation per qubit, gives $2 \times 6 \times 2 = 24$ trainable circuit parameters $\boldsymbol\theta$; the classical readout $W \in \mathbb{R}^{64\times 6}$ plus bias adds $384+64=448$; and the scalar gate $\sigma_{\text{res}}$ adds 1 -- a total of 473 trainable parameters for the $Z$-only quantum branch (1241 for QCR-XYZ, Section~\ref{sec:qcr-xyz}), several orders of magnitude smaller than the backbone or even the correlation module's own convolutions. This is worth stating plainly: whatever effect the QCR has (Section~\ref{sec:results}), it is not attributable to added classical model capacity, since the MLP-refiner control in Table~\ref{tab:main} is deliberately matched to this same, small parameter budget.

\begin{table*}[!t]
\centering
\small
\caption{Parameter count by component (analytically derived) and Fold 1 wall-clock training time} 
\label{tab:complexity}
\begin{tabular}{p{7.2cm}cc}
\toprule
Component & Trainable Parameters & Time/epoch (Fold 1) \\
\midrule
ResNet-50 backbone (frozen) & $\sim$23.5M (excluded) & -- \\
Correlation embedding (1$\times$1 convs, Layer-2/3 $\to$ 128-d) & impl.-dependent & -- \\
Correlation compression ($3{\times}3$ conv, $130\!\to\!64$ ch.) & $\approx 74{,}944$ (incl.\ bias) & -- \\
Decoder (transposed convs + skip fusion) & impl.-dependent & -- \\
\midrule
No refiner (full model, no QCR)         & $+0$    & $\approx$47s \\
Quantum QCR ($Z$-only: $24{+}448{+}1$)  & $+473$  & $\approx$90s \\
MLP-refiner (param-matched)             & $+473$  & $\approx$46s \\
Quantum QCR-XYZ ($24{+}1216{+}1$)       & $+1241$ & $\approx$146s \\
\bottomrule
\end{tabular}
\end{table*}

\begin{table*}[!t]
\centering
\small
\caption{Quantum-resource characterization for the evaluated configurations.}
\label{tab:quantum-resources}
\begin{tabular}{lcccccc}
\toprule
Variant & Qubits & Layers & Readout values & Trainable parameters & Time/epoch & Hardware depth / 2Q gates \\
\midrule
QCR ($Z$-only) & 6 & 2 & 6 & 473 & $\approx$90 s & 436 / 139$^{a}$ \\
QCR-XYZ & 6 & 2 & 18 & 1241 & $\approx$146 s & -- \\
MLP-refiner & -- & -- & -- & 473 & $\approx$46 s & -- \\
\bottomrule
\end{tabular}
\\[0.3em]

\end{table*}

The resource results highlight an important engineering tradeoff. QCR adds only 473 trainable parameters, but its measured training time is approximately twice that of the no-refiner and parameter-matched MLP conditions. QCR-XYZ increases the parameter count to 1241 and the measured training time to approximately 146 s per epoch, about 60\% above $Z$-only QCR, without producing a measurable accuracy benefit. The hardware characterization further shows that amplitude-embedding state preparation can lead to substantial circuit depth: the evaluated transpiled circuit reached depth 436 with 139 two-qubit gates per patch. These observations do not establish a general quantum-resource disadvantage because they concern one implementation and one backend, but they provide concrete resource measurements for the configuration studied here. QCR costs approximately twice the time of the no-refiner and MLP-refiner conditions, while QCR-XYZ adds a further approximately 60\% over QCR.

\subsection{Qualitative Results}
\label{sec:qualitative}

\begin{figure*}[!t]
    \centering
    % TODO: replace with the actual 1-shot prediction grid; this filename is a placeholder --
    % figure6_forest_deltas.pdf (used in Figure~\ref{fig:fold-comparison} above) was mistakenly
    % reused here and shows unrelated content (the delta-forest plot, not prediction masks)
    \includegraphics[width=\textwidth]{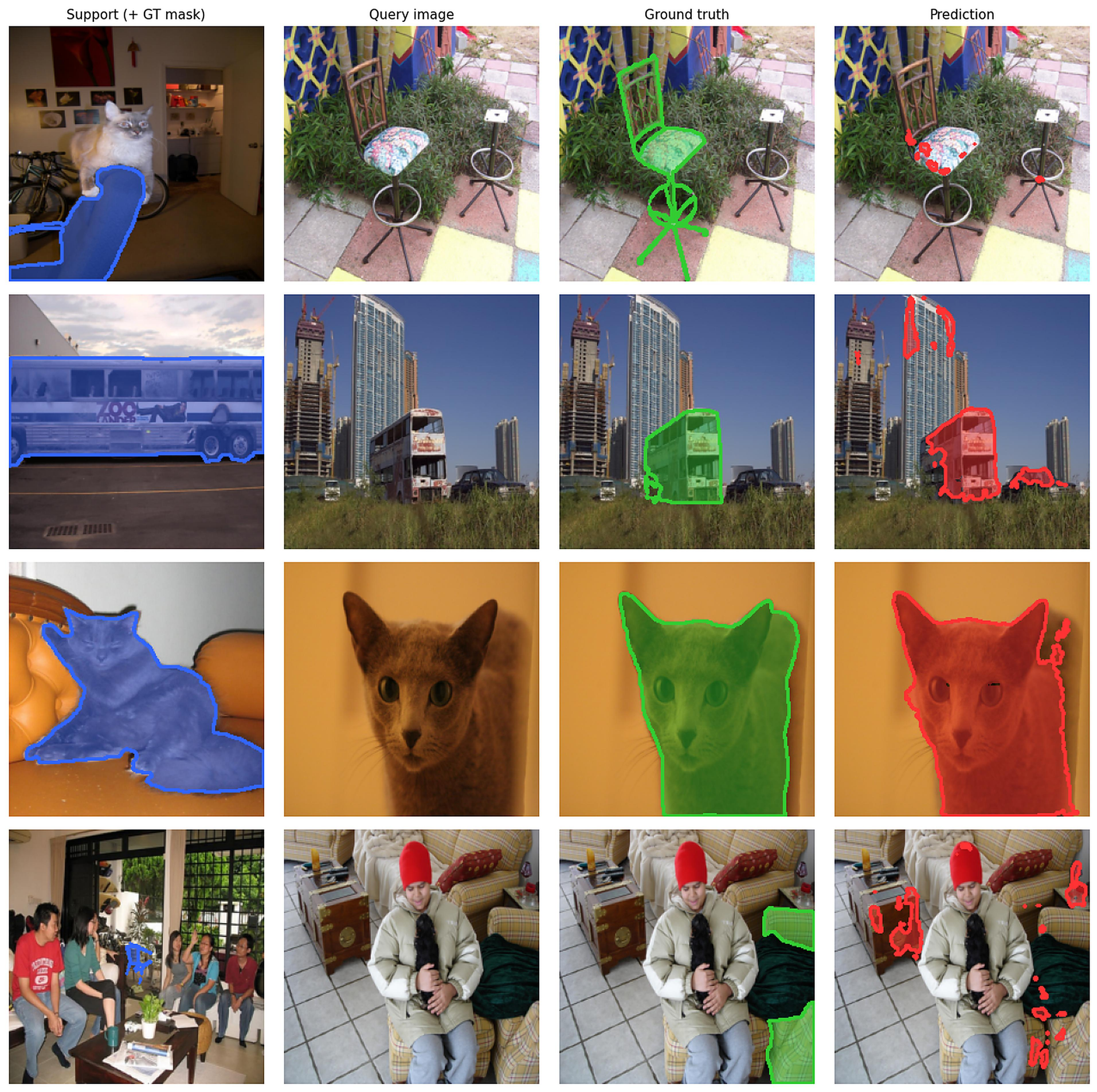}
    \caption{Representative 1-shot predictions from a quantum-augmented checkpoint (Fold 1). Columns, left to right: support image with ground-truth mask overlaid (blue), query image, query ground-truth mask overlaid (green), model prediction overlaid (red); mask boundaries are outlined for visibility against low-contrast backgrounds. Row 4's target class is \emph{chair} (a small fragment in the lower-right corner)fig; the model's prediction fails to localize it, illustrating sensitivity to small, low-contrast targets amid visually similar clutter.}
    \label{fig:qualitative}
\end{figure*}

Figure~\ref{fig:qualitative} shows representative 1-shot predictions selected to include both successful and unsuccessful cases rather than only favorable examples. Mask boundaries are rendered as solid outlines in addition to a semi-transparent fill, since fill alone can be difficult to distinguish against low-contrast backgrounds (see below). The first row (chair) shows a close match to ground truth, including correct rejection of visually similar distractor objects (the stools) in the background. The third row (cat) shows a generally accurate silhouette with a boundary error localized to the ear region, consistent with the resolution limitations discussed in Section~\ref{sec:limitations}. The second row (bus) illustrates a clearer failure mode: the predicted mask leaks into the background crane/building structure and shows an internal gap, plausibly reflecting a support-query appearance mismatch -- the support image is a plain, side-on view of a differently colored bus, while the query bus is viewed at an angle, partially occluded, and a different color. The fourth row is a genuinely instructive failure case once correctly read: the episode's target class is \emph{chair}, and the ground-truth mask is a small, partially-visible chair fragment in the lower-right corner of the query image, not the person seated centrally in it (an easy region to overlook without a contour outline, which is why we adopted one). The model fails to localize this small, low-contrast target and instead produces scattered false-positive predictions across visually similar upholstery and fabric elsewhere in the scene -- an example of the correlation module being drawn toward texture-similar clutter when the true target is small and awkwardly placed, rather than a data or visualization artifact.

\subsection{Error Analysis}
\label{sec:error-analysis}

Drawing on the four cases in Figure~\ref{fig:qualitative} and the broader pattern across validation episodes inspected during Section~\ref{sec:qualitative}, failures cluster around a small number of recognizable conditions rather than being uniformly distributed. Small or awkwardly-positioned targets are the clearest failure mode: the fourth row's chair fragment, occupying a small fraction of the image and abutting visually similar upholstery, is missed entirely in favor of texture-similar clutter elsewhere in the scene. Support-query appearance mismatch is the second recognizable failure mode: the second row's bus case shows the correlation module struggling when the support example's viewpoint, color, and occlusion state differ substantially from the query, leaking prediction mass into background structures with locally similar texture. By contrast, cases that succeed (rows 1 and 3) tend to share close support-query appearance and involve a target that is a large, contiguous, texturally distinct region -- conditions under which dense cosine-similarity correlation is expected to be most reliable in principle.

We do not yet have a quantitative, dataset-wide breakdown of error rate by object size, occlusion level, or support-query appearance similarity -- the observations above are drawn from inspection of a limited set of qualitative examples, not a systematic audit, and should be read as hypotheses about failure modes rather than measured error statistics. A quantitative version of this analysis (e.g., binning validation episodes by ground-truth mask area and reporting per-bin mIoU) is listed as future work (Section~\ref{sec:future-work}) rather than reported here, since we have not run it.

We also do not have evidence, from the data collected so far, that the \emph{quantum} refiner specifically changes which of these failure modes occur, as opposed to the classical baseline: the ablation study (Section~\ref{sec:ablation}) shows QCR, MLP-refiner, and noise-injection all producing statistically indistinguishable final mIoU, and we did not perform a qualitative side-by-side of classical vs.\ quantum predictions on the same failure cases shown in Figure~\ref{fig:qualitative}. Claiming that quantum refinement specifically helps with thin structures, boundary ambiguity, or background clutter -- beyond what an equivalently-sized classical perturbation achieves -- would not be supported by the evidence in this paper, and we do not make that claim.

\subsection{Hardware Validation}
\label{sec:hardware-validation}

All results reported use PennyLane's \texttt{default.qubit} simulator, both for training and for the accuracy comparison in Section~\ref{sec:results}. That leaves a separate, narrower question open: does a \emph{trained} QCR circuit compute consistently when executed on a real quantum device, rather than only on the noiseless simulator it was optimized against? This subsection reports a single-checkpoint check of that question. It does not re-evaluate, and is not a substitute for, the mIoU comparison above -- the null accuracy result in Section~\ref{sec:results} holds regardless of what this subsection finds, since it concerns numerical fidelity of the circuit's output, not downstream segmentation performance.

 We took the Fold 1, seed-42 $Z$-only QCR checkpoint (validation mIoU $0.6087$, matching Table~\ref{tab:per-seed}; final $\sigma_{\text{res}}=0.1253$, within the Fold 1 quantum gate range reported in Section~\ref{sec:gate-behavior}) and sampled 18 correlation-map patches ($64$-dimensional, unit-normalized) from genuine Fold 1 validation episodes, hooked directly from the tensor entering \texttt{qrefine} in the trained model's forward pass. We built an equivalent Qiskit~\cite{Qiskit2019} circuit -- identical ansatz (amplitude embedding, two $R_Y$/$R_Z$ layers, ring-CNOT entanglement, single-qubit $\langle Z\rangle$ readout) -- and executed it on IBM's \texttt{ibm\_fez} (156 qubits) via Qiskit Runtime's \texttt{EstimatorV2} primitive, resilience level 1, 4096 shots per observable. Transpiled circuits reached depth 436 with 139 two-qubit gates per patch on the backend's native \texttt{sx}/\texttt{rz}/\texttt{cz} basis -- a non-trivial state-preparation cost from \texttt{initialize}-based amplitude embedding, consistent with the expressivity/cost discussion in Section~\ref{sec:method}, and one reason we scoped this check to a small patch sample rather than the full validation set.

\begin{figure*}[!t]
    \centering
    \includegraphics[width=0.95\textwidth]{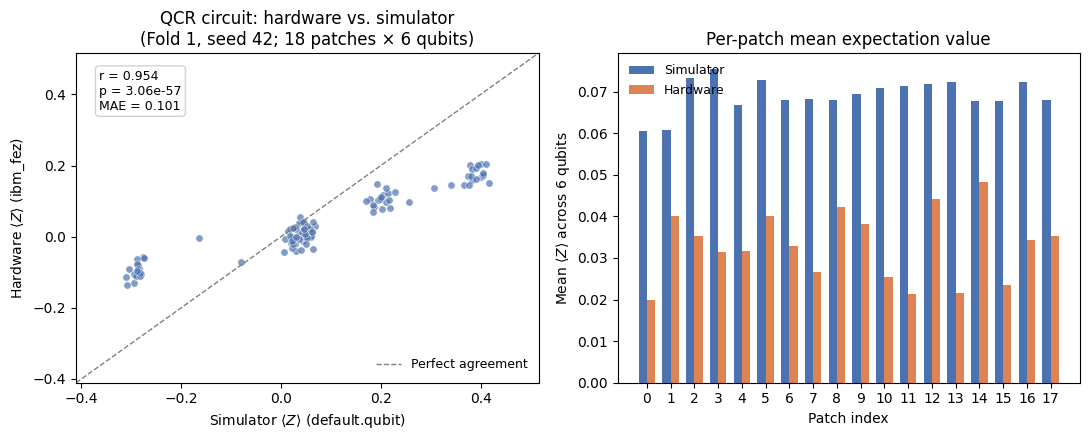}
    \caption{QCR hardware (\texttt{ibm\_fez}) vs.\ simulator (\texttt{default.qubit}) agreement, Fold 1 seed-42 checkpoint, 18 validation patches $\times$ 6 qubits,. \textbf{Left:} per-observable scatter of hardware $\langle Z\rangle$ against simulator $\langle Z\rangle$ across all 108 (patch, qubit) pairs\textbf{Right:} per-patch mean $\langle Z\rangle$ (averaged across the 6 qubits), simulator vs.\ hardware side by side.}
    \label{fig:hardware-validation}
\end{figure*}

\textbf{Validation of the simulator-to-hardware mapping.} Before the reported hardware measurement, the PennyLane-to-Qiskit implementation was checked against the noiseless simulator to verify observable ordering and amplitude-vector bit ordering. This check identified a convention mismatch between the two implementations. After correcting the mapping, the hardware circuit agreed closely with the simulator baseline: $r=0.954$ ($p=3.1\times10^{-57}$) and MAE $=0.101$ across 108 observable values from 18 validation patches (Figure~\ref{fig:hardware-validation}). We report the corrected result only; the purpose of the check is to establish circuit-level fidelity, not to quantify a hardware accuracy advantage.

\textbf{Scope.} This result covers a single checkpoint (Fold 1, seed 42, $Z$-only QCR) and a small patch sample ($n=18$ patches, $108$ expectation values); it does not extend to other folds, seeds, the MLP-refiner or QCR-XYZ conditions, or the full 1000-episode validation set, none of which were run on hardware here.

\section{Discussion}
\label{sec:discussion}

The pooled null result ($\Delta = +0.0001$ mIoU, $t(11)=0.10$, $p=0.92$, Cohen's $d_z=0.03$, across all 12 fold$\times$seed pairs) is indistinguishable from the MLP-refiner and QCR-XYZ controls (Section~\ref{sec:ablation}). The question this section addresses is not why a small positive effect appears despite a non-saturating gate, but whether any real effect exists -- and the verified evidence says no. We organize the discussion around the following points.

\begin{enumerate}

\item A small trained residual at this location does not measurably help or hurt, regardless of what computes it. This holds across all four folds. Quantum QCR, MLP-refiner, and QCR-XYZ land within a 0.0029 mIoU band of the baseline and of each other; pooled quantum-vs-classical difference is $+0.0001$. MLP-refiner shows $+0.0003$ from classical ($p=0.93$) and $-0.0002$ from quantum ($p=0.95$); QCR-XYZ shows $+0.0005$ from classical ($p=0.81$); noise-injection is null too ($+0.0003$, $p=0.78$). Fold 1's quantum decrease is consistent across its 3 seeds, but this does not generalize -- three of four folds show inconsistent per-seed signs, and the one positive fold (Fold 3) is driven mostly by a single seed. MLP-refiner and QCR-XYZ tracking quantum's per-fold sign across all folds (Tables~\ref{tab:main},~\ref{tab:xyz-allfolds}) argues against a quantum-specific mechanism.

\item PQC trainability limitations, and what QCR-XYZ does and doesn't tell us. QCR-XYZ tested the measurement-dimensionality bottleneck by tripling readout degrees of freedom (Section~\ref{sec:methodology-remark-on-bottleneck}) and produced no change in either direction, weakening this as a complete explanation. Ansatz depth, embedding scheme, and qubit count remain untested (Section~\ref{sec:future-work}). Given the null spans all four folds, this question is secondary to Point 1: even a perfectly-trained PQC needs a correction worth finding.

\item Per-fold heterogeneity is resolved as noise, not signal.Variation shows no reproducible relationship to fold identity: Fold 1's negative effect is consistent across seeds, Fold 3's positive mean is driven by one seed, and remaining folds are near zero. Given the near-zero pooled estimate and similar patterns across controls, we read this as sampling and training variability.

\item The gate's failure to saturate remains an open question independent of effect sign. A gate plateauing near its $0.1$ initialization is consistent with either a bounded, small-magnitude correction or a near-uninformative gradient signal. Verified per-epoch trajectories (Section~\ref{sec:gate-behavior}) show quantum, MLP-refiner, and QCR-XYZ in a comparable band with no consistent separation -- more consistent with the residual-gating scheme itself than a property unique to variational circuits.

\item \{Barren plateaus remain a partial, but weakened, explanation. Our circuit is shallow (2 layers, 6 qubits), on the favorable end of the depth/width regime~\cite{Cerezo2021}. Amplitude-embedding 64 dimensions into 6 qubits could produce weak gradients, but QCR-XYZ's null result on widened readout points more toward the embedding side, ansatz expressivity, or genuine absence of signal than toward measurement readout specifically. 
\item The module's small parameter count is consistent with a negligible effect. QCR is parametrically tiny relative to the architecture (473--1241 parameters vs.\ a $\sim$23.5M-parameter frozen backbone, Section~\ref{sec:complexity}). 

\item Four compounding factors underlie QCR's parity with its classical counterpart:
\begin{enumerate}
    \item \emph{Task granularity} -- prior positive PQC-vision results largely target image-level classification~\cite{Henderson2020Quanvolutional,Mari2020TransferLearning}, whereas dense segmentation requires per-location correction composing into coherent boundaries, a harder regime for a 6-qubit circuit.
    \item \emph{Position in the pipeline} -- in prior work the PQC often \emph{is} the classifier~\cite{Cong2019QCNN,SchuldBocharov2020CircuitCentric}; here QCR is a residual add-on after a ResNet-50 backbone and substantial decoder stack (Section~\ref{sec:complexity}), leaving little room to matter.
    \item \emph{Narrow measurement channel} -- only 6 (or 18, for QCR-XYZ) real values correct a 64-dimensional feature per location (Section~\ref{sec:methodology-remark-on-bottleneck}); QCR-XYZ's null result on a widened channel suggests the first two factors dominate over circuit expressivity.
    \item \emph{Methodological comparison} -- much prior positive literature lacks a parameter-matched control, multiple seeds, or log-level verification, the three checks applied here that reversed our own first-pass result from significant to null (Section~\ref{sec:results}).
\end{enumerate}

\item Hardware fidelity is separate from accuracy and should not be conflated. The hardware check ($r=0.954$ against simulator baseline, Section~\ref{sec:hardware-validation}) confirms only that the trained circuit survives execution on real hardware; it cannot bear on accuracy, which the pooled null already settles independently. High fidelity is not evidence for usefulness, nor low fidelity evidence against it. 
\end{enumerate}
\section{Limitations}
\label{sec:limitations}

The findings should be interpreted within the scope of the evaluated PASCAL-5\textsuperscript{i} benchmark, six-qubit two-layer circuit, and experimental configuration. Results may differ for other datasets, architectures, circuit designs, or resource regimes. The hardware experiment provides a limited validation of circuit behavior rather than a comprehensive hardware performance assessment. In addition, the study does not establish quantum computational or runtime advantage; it evaluates whether the investigated quantum refinement produces a measurable segmentation benefit under controlled, parameter-matched conditions. These considerations limit generalization but do not alter the main finding that the evaluated quantum module provides no statistically significant improvement over the corresponding classical alternatives.

\section{Future Work}
\label{sec:future-work}

Several directions follow naturally from the controlled null result. First, direct measurements of PQC gradient norms during training would help distinguish limited trainability from a genuine absence of useful refinement signal. Second, isolated profiling should measure FLOPs, peak GPU memory, and inference latency under controlled hardware conditions, complementing the training-time measurements reported here. Third, systematic sweeps of qubit count, circuit depth, and data-encoding strategy would establish whether the observed result is specific to the six-qubit, two-layer amplitude-encoded configuration or persists across a broader design space. Fourth, repeating the study with a fine-tuned backbone, higher input resolution, additional FSS architectures, and datasets such as COCO-20$^i$~\cite{Nguyen2019COCO20i} would test external validity. Finally, broader hardware evaluation across checkpoints and devices, together with hardware-aware training or realistic noise models, would provide a stronger assessment of deployment-level behavior.

These extensions are deliberately presented as future experiments rather than completed evidence. No claim is made here about untested qubit counts, circuit depths, hardware generalization, or quantum advantage beyond the configuration and validation scope reported in this paper.

\section{Conclusion}
\label{sec:conclusion}

This study evaluated whether a small variational quantum circuit can provide measurable value when inserted as a correlation-refinement module in a classical few-shot semantic segmentation pipeline. Under the four-fold PASCAL-5$^i$ protocol with three seeds per fold, the six-qubit QCR module produced a mean mIoU difference of only $+0.0001$ relative to the classical baseline, with no statistically significant difference ($t(11)=0.10$, $p=0.92$, $d_z=0.03$). The parameter-matched MLP refiner and expanded-measurement QCR-XYZ variant produced similarly small, non-significant differences. The evidence therefore does not support a quantum-specific accuracy benefit for the evaluated configuration.

The result is informative because the comparison was designed to separate the effect of quantum computation from the generic effect of adding a small learned residual transformation. The matched controls, multiple seeds, fold-wise analysis, and verification against training logs all point to the same conclusion: within this architecture and resource regime, the quantum refiner behaves much like a small additional transformation whose contribution to final segmentation accuracy is negligible. We do not interpret this as evidence that quantum methods cannot benefit few-shot segmentation more generally; rather, it establishes a controlled negative result for the specific design studied here.

The resource analysis also illustrates an important engineering consideration. QCR adds only 473 trainable parameters but approximately doubles measured training time relative to the no-refiner and MLP conditions, while QCR-XYZ increases training time by approximately 60\% beyond $Z$-only QCR without improving accuracy. The hardware experiment provides a separate validation: for one trained checkpoint and 18 validation patches, the corrected circuit output agreed closely with the noiseless simulator ($r=0.954$, MAE $=0.101$). This result concerns circuit fidelity, not segmentation performance.

Taken together, the study provides a reproducible benchmark for evaluating small quantum components inside classical vision systems. The main lesson is methodological: claims about quantum benefit should be supported by parameter-matched classical controls, repeated seeds, transparent statistical analysis, resource measurements, and, where hardware execution is relevant, direct simulator-to-device validation. Under those criteria, the QCR configuration evaluated here does not demonstrate a measurable task-level advantage.

\section*{Reproducibility and Data Availability}
\label{sec:reproducibility}

\textbf{Reproducibility.} Comparisons are reproducible from the supplied implementation, training logs, and figure-generation scripts, using the publicly available PASCAL VOC 2012 dataset {URL :{\url{http://host.robots.ox.ac.uk/pascal/VOC/voc2012/}} under the standard PASCAL-5\textit{i} split ~\cite{Shaban2017OSLSM}}.

\textbf{Training configuration.} Input resolution 224$\times$224; AdamW optimizer, base learning rate $3\times10^{-4}$ (classical/PQC parameters) and $9\times10^{-6}$ (backbone, once unfrozen); loss is BCE plus Tversky ($\alpha=\beta=0.5$); batch size 16 episodes; backbone frozen for the reported 15-epoch runs; validation uses a fixed, seeded set of 1000 episodes per fold; seeds $\{42, 43, 44\}$. QCR uses 6 qubits, 2 variational layers, amplitude embedding, $Z$-expectation readout, and a residual gate initialized to 0.1.

\textbf{Hardware validation.} Evaluated checkpoint: Fold-1, seed-42 QCR (best epoch 12/15, validation mIoU 0.6087). Backend: IBM \texttt{ibm\_fez} (156 qubits), selected via \texttt{QiskitRuntimeService.least\_busy}. Execution used Qiskit Runtime EstimatorV2, resilience level 1, 4096 shots/observable. Transpiled circuit: \texttt{sx/rz/cz} basis, depth 436, 139 two-qubit CZ gates per patch. Corrected result (18 validation patches): $r=0.954$, MAE $=0.101$.

\
\section*{Acknowledgment}

H. Shakir conceived the study, designed the methodology, implemented the classical and hybrid quantum--classical models, conducted the experiments and statistical analyses, performed the hardware validation, and wrote and revised the manuscript. The author declares no competing interests.
\\
\appendix
\section{Detailed Algorithms}
\label{app:algorithms}
\begin{algorithm}[htbp]
\caption{QOSS-Corr Forward Pass and Loss Computation (single episode)}
\label{alg:qoss-corr}
\begin{algorithmic}[1]
\Require Support image $I_s$, support mask $M$, query image $I_q$, ground-truth query mask $y$
\Require Pretrained ResNet-50 backbone $\phi(\cdot)$; correlation fusion weights $W_{\text{fuse}}$ (per scale); QCR parameters $\boldsymbol\theta$, projection $W_{\text{out}}$, residual gate $\sigma_{\text{res}}$ (per scale); decoder weights
\Ensure Predicted segmentation mask $\hat{y} \in [0,1]^{224\times224}$; loss $\mathcal{L}$

\State $\mathbf{s}^{(1)}, \mathbf{s}^{(2)}, \mathbf{s}^{(3)} \gets \phi(I_s)$ \Comment{layer1/2/3 support features}
\State $\mathbf{q}^{(1)}, \mathbf{q}^{(2)}, \mathbf{q}^{(3)} \gets \phi(I_q)$ \Comment{layer1/2/3 query features}
\State $\mathbf{s}^{(1)}_{\text{skip}} \gets \text{Project}_{32}(\mathbf{s}^{(1)})$ \Comment{decoder skip connection, Sec.~\ref{sec:method-backbone}}
\For{scale $k \in \{2, 3\}$} \Comment{layer2, layer3}
    \State $\mathcal{F} \gets \{(i',j') : M^{(k)}_{i',j'} = 1\}$ \Comment{mask resized to scale $k$}
    \State compute $\hat{\mathbf{s}}^{(k)}_{i,j}$ via Eqs.~\eqref{eq:cosine-sim}--\eqref{eq:attended-support} \Comment{masked cosine attention}
    \State compute $m^{(k)}_{i,j}, x^{(k)}_{i,j}$ via Eq.~\eqref{eq:mean-max-sim} \Comment{mean/max similarity stats}
    \State $\mathbf{c}^{(k)} \gets W_{\text{fuse}}^{(k)} * \big[\mathbf{q}^{(k)} \Vert \hat{\mathbf{s}}^{(k)} \Vert m^{(k)} \Vert x^{(k)}\big]$ \Comment{Eq.~\eqref{eq:correlation-fusion}}
    \For{each spatial location $(i,j)$}
        \State $|\psi\rangle \gets \text{AmplitudeEmbed}\big(\mathbf{c}^{(k)}_{i,j} / \lVert \mathbf{c}^{(k)}_{i,j} \rVert\big)$ \Comment{normalize, 6-qubit embedding}
        \State $|\psi'\rangle \gets U(\boldsymbol\theta)\, |\psi\rangle$ \Comment{$L{=}2$ entangling layers, Sec.~\ref{sec:method-quantum}}
        \State $\langle Z \rangle \gets \big(\langle\psi'|Z_0|\psi'\rangle, \dots, \langle\psi'|Z_{n-1}|\psi'\rangle\big)$
        \State $\mathbf{c}'^{(k)}_{i,j} \gets \mathbf{c}^{(k)}_{i,j} + \sigma_{\text{res}}^{(k)} \cdot W_{\text{out}}^{(k)}(\langle Z \rangle)$ \Comment{gated residual}
    \EndFor
\EndFor
\State $\mathbf{c}_{\text{fused}} \gets \text{Conv}_{3\times3}\big[\text{Upsample}(\mathbf{c}'^{(2)}) \Vert \text{Upsample}(\mathbf{c}'^{(3)})\big]$ \Comment{multi-scale fusion, Sec.~\ref{sec:method-decoder}}
\State $\hat{y} \gets \text{Decoder}\big(\mathbf{c}_{\text{fused}}, \mathbf{s}^{(1)}_{\text{skip}}\big)$ \Comment{transposed-conv upsampling to $224{\times}224$}
\State $\mathcal{L} \gets \mathcal{L}_{\text{BCE}}(\hat{y}, y) + \mathcal{L}_{\text{Tversky}}(\hat{y}, y)$ \Comment{Eqs.~\eqref{eq:bce-loss}--\eqref{eq:total-loss}}
\State \Return $\hat{y}, \mathcal{L}$
\end{algorithmic}
\end{algorithm}

\begin{algorithm}[htbp]
\caption{QOSS-Corr Training Loop (one fold $\times$ seed $\times$ condition)}
\label{alg:training-loop}
\begin{algorithmic}[1]
\Require Training episodes drawn from 15 classes (three folds excluding the held-out fold); fixed, seeded validation set of 1000 episodes for the held-out fold
\Require Max epochs $E{=}15$; early-stopping patience $P{=}10$; backbone freeze duration $E_{\text{freeze}}{=}15$ epochs
\Require Pretrained, initially-frozen ResNet-50 backbone; randomly initialized correlation, decoder, and (if condition $\ne$ classical) refiner weights
\Ensure Best checkpoint by validation mIoU

\State Initialize AdamW with parameter groups: classical params at lr $3{\times}10^{-4}$; if condition $\ne$ classical, refiner params at lr $3{\times}10^{-4}$ \Comment{backbone group added only after unfreezing}
\State $\text{best\_mIoU} \gets -\infty$; $\text{no\_improve} \gets 0$
\For{epoch $= 1$ \textbf{to} $E$}
    \If{epoch $= E_{\text{freeze}} + 1$} \Comment{not reached by any run reported here, since $E = E_{\text{freeze}} = 15$}
        \State Unfreeze backbone; add backbone parameter group at lr $9{\times}10^{-6}$
    \EndIf
    \State model.train()
    \For{each training episode (support image/mask, query image/mask) sampled from training classes}
        \State $\hat{y}, \mathcal{L} \gets$ Algorithm~\ref{alg:qoss-corr} forward pass (support/query augmented independently: scale jitter, flip, color jitter)
        \State $\mathcal{L}$.backward(); optimizer.step(); optimizer.zero\_grad()
        \State accumulate running TrainLoss and TrainmIoU (per-class intersection/union) over the epoch
    \EndFor
    \State model.eval()
    \State \textbf{with} torch.no\_grad(): evaluate on the fixed 1000-episode validation set (flip test-time augmentation) $\to$ ValLoss, ValmIoU \Comment{per-class accumulated mIoU, Sec.~\ref{sec:setup}}
    \If{ValmIoU $>$ best\_mIoU}
        \State $\text{best\_mIoU} \gets$ ValmIoU; save checkpoint; $\text{no\_improve} \gets 0$
    \Else
        \State $\text{no\_improve} \gets \text{no\_improve} + 1$
    \EndIf
    \If{$\text{no\_improve} \ge P$}
        \State \textbf{break} \Comment{early stopping; not triggered by any run reported here within $E=15$ epochs}
    \EndIf
\EndFor
\State \Return checkpoint with highest recorded ValmIoU
\end{algorithmic}
\end{algorithm}

\bibliographystyle{IEEEtran}
\bibliography{references}

\end{document}